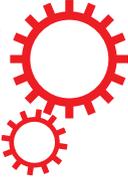

OPEN

# Evaluation of Direct Haptic 4D Volume Rendering of Partially Segmented Data for Liver Puncture Simulation

Andre Mastmeyer, Dirk Fortmeier & Heinz Handels

This work presents an evaluation study using a force feedback evaluation framework for a novel direct needle force volume rendering concept in the context of liver puncture simulation. PTC/PTCD puncture interventions targeting the bile ducts have been selected to illustrate this concept. The haptic algorithms of the simulator system are based on (1) partially segmented patient image data and (2) a non-linear spring model effective at organ borders. The primary aim is to quantitatively evaluate force errors caused by our patient modeling approach, in comparison to haptic force output obtained from using gold-standard, completely manually-segmented data. The evaluation of the force algorithms compared to a force output from fully manually segmented gold-standard patient models, yields a low mean of 0.12 N root mean squared force error and up to 1.6 N for systematic maximum absolute errors. Force errors were evaluated on 31,222 preplanned test paths from 10 patients. Only twelve percent of the emitted forces along these paths were affected by errors. This is the first study evaluating haptic algorithms with deformable virtual patients *in silico*. We prove haptic rendering plausibility on a very high number of test paths. Important errors are below just noticeable differences for the hand-arm system.

Virtual reality (VR) surgery simulation with needle insertion into blood vessels[1, 2] or for liver biopsy[3, 4] is a current field of research, which deals with the topics of visual and haptic rendering (visuo-haptics) as well as the generation of virtual patient models. A major part of such simulation systems, namely the force feedback during needle insertion, has been surveyed[5, 6]. The downside of most systems is their inability to easily simulate an intervention on new patient data (patient-specific) with little manual segmentation preparation time and effort, for which an overview was recently given by ref. 7. Typical FEM-based simulators often need suitable volumetric meshing and on-line locally-adaptive remeshing of the organ models[8]. In addition, these simulators usually require a fully segmented 3D patient, which leads to high segmentation efforts and avoids the use of VR simulators in a clinical setting. This fact and missing patient motion dynamics are the major drawback of state of the art simulators. Our works using 3D and 4D CT image data[9, 10] remedy this situation.

In ref. 11, a 4D simulator for percutaneous transhepatic cholangiography (PTC) is presented. There, surface mesh models have to be created and following this paradigm neither direct haptic nor visual rendering of patient image data is used. A similar procedure, percutaneous transhepatic-cholangiodrainage (PTCD), is a needle insertion intervention in which dilated bile ducts, caused by cholestasis, are punctured, (see Fig. 1) to relieve the patient by drainage via 17 G catheters[12]. Cholestasis can be caused by i.e. gallstones or tumors in the common bile duct (CHD). To reach the target (right hepatic bile duct, RHD), the needle has to be inserted between the ribs (intercostal spaces) and into the liver. Modern intervention techniques use an ultrasound (US) probe with an attached needle guide (Fig. 2) that helps to keep the needle on paths inside the US image plane. For improved planning, a CT volume image can be acquired before performing the intervention.

Previously, we have presented the predecessor AcusVR-3D[13] featuring fully manually-segmented patient models and our new simulator AcusVR-4D[9] with simulated US imaging needle guidance as a valuable training and planning tool. The old system consisted of a Geomagic Phantom Premium 1.5 6DOF haptic device with a combination of shutter glasses and a CRT monitor for VR immersion. This 3D VR simulator was mainly aimed

Institute of Medical Informatics, University of Luebeck, Luebeck, 23552, Germany. Correspondence and requests for materials should be addressed to A.M. (email: mastmeyer@imi.uni-luebeck.de)

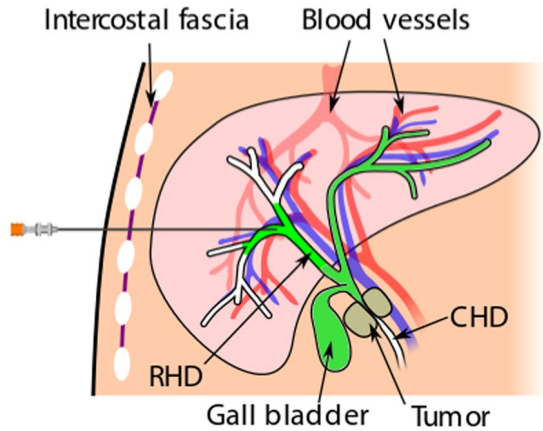

**Figure 1.** Schematic overview of needle insertion into the right hepatic duct (RHD): A tumor closes the common hepatic duct (CHD) and cholestasis results.

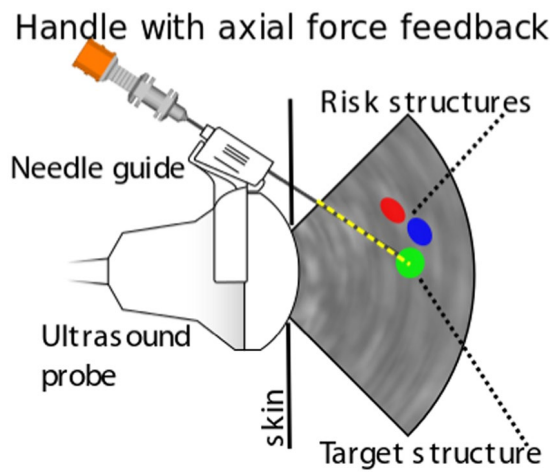

**Figure 2.** Schematic overview of ultrasound (US) guided needle insertion into the right hepatic duct (RHD): A US probe is put on the skin and captures a dilated hepatic duct (target structure). The needle is guided on a rail at the side of the probe. Puncturing of risk structures such as blood vessels should be avoided.

at the training of lumbar punctures. A user study was conducted and training has been shown to be effective[13]. However, AcusVR-3D and other state of the art systems rely on time-consuming complete manual segmentations. Depending on the anatomical site, standard technique manual segmentations can take more than 40 hours[14] for a single patient data set.

Regarding haptics, needle insertion procedure simulations must mimic stiffness, cutting and friction forces at the needle tip and shaft[5, 6, 15]. Phases during needle insertion to be modeled are: pre-puncture pullback, puncture incident, penetration phase and tissue exit. In this paper, we focus on needle insertion and the axial force components. Typically, the retraction of the needle emits the same shaft forces as encountered during insertion without the cuts from the needle tip. Needle deflection (bending), tissue deformation and non-axial forces[10, 16] are also relevant for realistic, real-time, visuo-haptic simulations. However, these topics are beyond the scope of this work.

Currently, haptic rendering in these visuo-haptic simulations raises the need for evaluation methods and detailed studies of the force output at the handle regarding needle insertion (Fig. 2). A preliminary study using one static 3D patient was presented in ref. 17. In this paper, we focus the quantitative haptic evaluation of the axial force output of our new simulator AcusVR-4D and generate a high number of test paths from ten test patients for the simulation of PTCD. Along these test paths, forces from reference and new haptic algorithms are calculated and compared. In the absence of *in vitro/vivo* gold-standard force measurements, we use reference forces based on previously evaluated fully segmented patient data (*in silico*)[9, 13]. Qualitatively, the haptic parameters determining the force output using our new system regarding liver structures, were tuned by two medical experts experienced in PTCD liver punctures[9]. Here, we present the quantitative haptics evaluation counterpart. We think the design of the presented evaluation experiment is a practicable way to bridge the gap between qualitative and still missing quantitative evaluations of virtual haptic simulations.

The aim of this study is to compare axial force errors of simulated needle insertion force feedback of a full patient segmentation, against a partial segmentation based force calculation algorithm. We also characterize the differences in terms of the magnitude and location of the errors found. The errors observed here are the quantitative marks of our virtual training simulator based on our virtual patient modeling approach, where only certain key structures need reviewed segmentations.

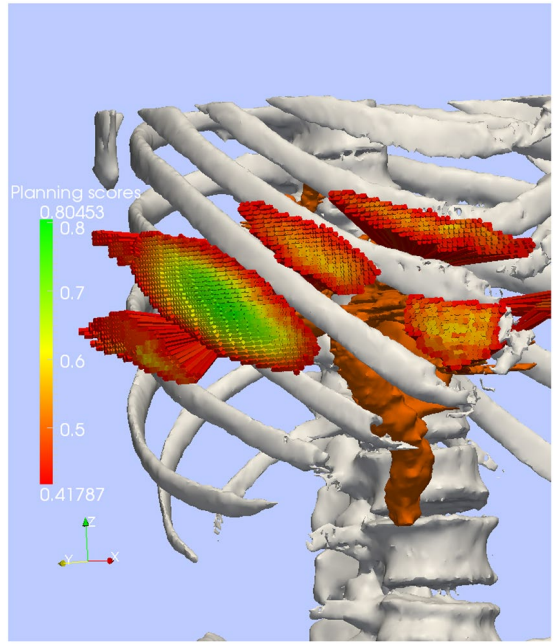

**Figure 3.** Reference paths (colored), bile ducts (brown) and bone (white). On our paths reaching the bile ducts scoring reveals best paths in the central green area mainly between the 6th and 7th rib. Path coloring corresponds to the quality of the path deduced from the soft constraints.

The article is organized as follows: First, quantitative results using the evaluation methods on page 12 of our haptic sub-system for PTC(D) are given in section "Results" on the current page and discussed on the following page. The methods on page 6 start with an overview of AcusVR-4D and its hardware and software components. Then, the concept of our virtual patient modeling is summarized on page 7. Details of our method for the haptic simulation of axial forces during needle insertion, are presented in the following section dealing with haptic rendering on page 9. The evaluation methods section on page 12, shows our general framework for evaluation and interpretation of our direct needle force rendering quality.

## Results

We applied our methods to 10 patient (#1-#10) clinical CT data sets and compared the new haptics method using "partial segmentation masks augmented by our patient-specific transfer functions," against the haptic algorithm based on gold-standard fully manually-segmented data.

The benefit of our patient modeling[18] vs. manually driven full volume segmentation, is the significantly shorter time frame (<10x) needed to provide a virtual patient model. This is to a large extent achieved by using the threshold-based ($t_0 \ldots t_2$) transfer functions dependent on the voxel position[18, 19]. The transfer functions cover the voluminous, but less relevant and easily segmentable tissues, such as skin, fat, bone and air cavities inside the body. In this study, we introduce and report (1) segmentation errors using new local metrics, i.e. well-known measures are evaluated only on the planned puncture paths, rather than the full volume; (2) we show the axial force errors caused by these segmentation errors.

Regarding haptics, this is the first study in which a direct haptic volume rendering method is tested on 10 patients and 31,222 test paths (i.e. ca. 3000 paths per patient). Exemplary test path planning results can be seen in Fig. 3. High quality paths are colored in green.

We report the segmentation results, shown in Figs 4 and 5, in terms of mean surface distance (MSD) and Hausdorff distances (HSD), evaluated on the individual test paths. Using semi-automatic segmentation methods described in refs 18, 20, 21, the depth difference errors, which correspond on average to mean surface distances (MSD), are lower than the acceptable 2 mm (Fig. 4). Regarding HSD to characterize outliers, we encounter errors exeeding 3 mm for the liver and 2 mm for the soft tissue (Fig. 5). The important target structures, bile ducts, fall below the lower bound in both metrics.

In Figs 6 and 7, the force errors are shown as box plots, with root mean squared or maximum absolute errors (RMSE, MAE) in the left or right column. In Fig. 6, we can observe RMS errors lower than max. 0.25 N and MAE errors less than 1.0 N, except for some displaced outliers around 1.5 N. Systematic maximal force error medians of 0.7 N and rare sporadic peak errors, around 1.4 N occur in some patients (Fig. 7). All systematic median MA errors are less than or equal to 0.7 N and 88% of the occuring forces are, on average, identical (Fig. 8). The total average over 31,222 paths from ten patients shows RMSE or MAE of $0.12 \pm 04\,N$ with 0.05% number of top outliers - or $0.65 \pm 0.18\,N$ with 3.73% top outliers.

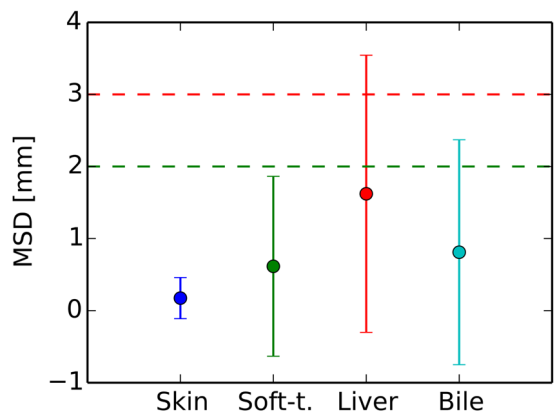

**Figure 4.** Mean and standard deviations of per path mean surface distances (MSD). Skin tissue depth lags are best. A depth lag of 2 mm (dashed green line) is considered a low just noticeable spatial error for salient needle force events to occur.

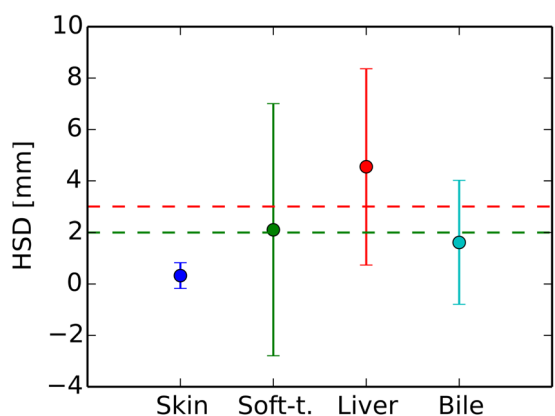

**Figure 5.** Mean and standard deviations of per path Hausdorff surface distances (HSD). Skin tissue depth lags are best. A depth lag of 3 mm (dashed red line) is considered a just noticeable spatial error for salient needle force events to occur.

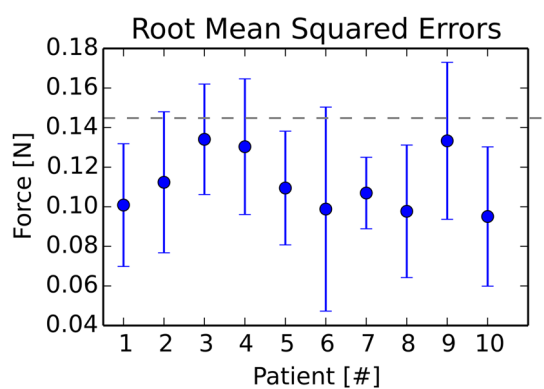

**Figure 6.** Change of segmentation model: Force metric RMSE per path.

Mean force errors are, for the most part, below the just noticeable difference (JND) force threshold, elaborated on page 13, of 0.145 N (Fig. 6).

To sum up, the spatial distance errors (MSD) are below just noticeable thresholds, except for some (HSD) outliers of the liver. While small, almost imperceptible, RMS force errors occur, a systematic bias of about 0.7 $N$ on average, can be observed in terms of MAE. These and the outliers of ca. 1.4 $N$ prominent in some patients for the change of segmentation experiment, are clarified in the discussion. In total, we achieve on average 88% of exactly identical emitted forces (Fig. 8), i.e. only 12% of forces are affected by errors of any magnitude.

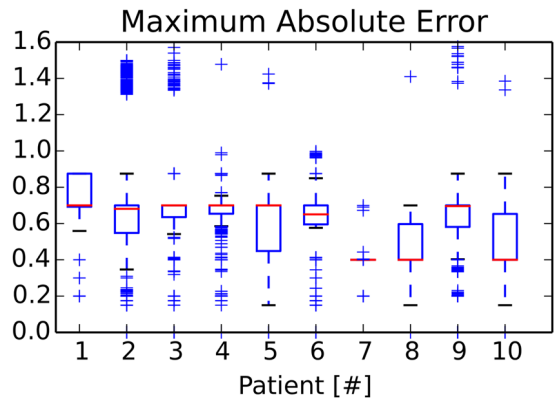

**Figure 7.** Change of segmentation model: Force metric MAE per path.

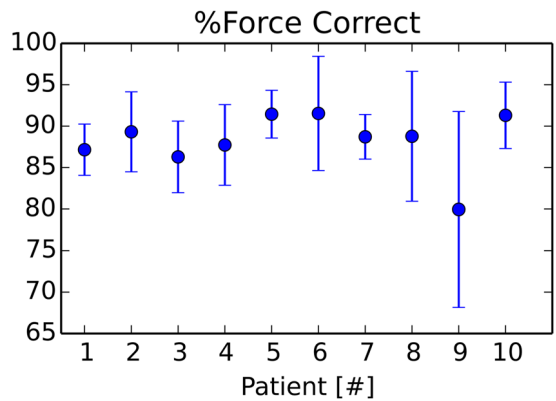

**Figure 8.** Change of segmentation model: Error bars comparing the percentage of exactly identical force outputs per path for each patient.

### Discussion
Our qualitatively accepted VR system, AcusVR-4D[9] for needle insertion simulation, has been evaluated quantitatively in terms of haptic rendering plausibility, with positive results for the influence of usual segmentation errors. The conducted "change of segmentation model" experiment (Figs 6 and 7) validates force output from the use of semi-automatic segmentation masks augmented by the transfer function classifier[9, 18], with promising results. Our haptic evaluation framework, study design and interpretation guide-lines, using state of the art spatial and force JNDs, could be readily applicable to other simulators. However, the evaluation framework and the presented results are simulation-software-based (*in silico*) in absence of *in vivo* and *in vitro* reference measurement data, which are also difficult to supply. Needle steering can not be assessed by axial forces and deformations[22–24] alone, and non-axial needle-tissue interaction forces are found along the entire embedded needle length. However, in our situation (axial punctures) the forces from breathing motion are mainly directed orthogonally to the needle insertion direction, so that they do not contribute saliently during needle advance. Thus, regarding axial force we evaluate the relevant haptic aspect of the simulation compared to gold-standard virtual patients and reference force output approved by medical experts.

With regards to skin, soft-tissue and bile segmentation, the spatial errors (MSD, HSD) are small. We often observed that the manual skin segmentation is detected slightly later on a needle path than the skin found by the threshold. The resulting systematic maximum force error of 0.7 N at the skin surface is not severe in our context, as the needle enters the body through a small ski n incision made by the surgeon.

In terms of quantitative force comparison, the errors are marginal and the number of outliers is small. Their values are negligible in the RMS metric. Maximal errors of 1.6 N in the MAE metric can be located, explained and well justified by the following: (1) at the skin a maximum 0.7 N difference is calculated and (2) immediately above the fascial layer, a somewhat differently classified voxel (group) can cause phase 1 (Pre-X) to begin from a different force level resulting in misaligned puncture peaks (cf. Sec. on page 9). This results in ca. 1.6 N maximal error. Thus, for the change of the segmentation model, errors are caused by slight differences in the manual expert segmentations and the segmentations resulting from our patient modeling approach. As long as the sequence of events is guaranteed to be the same, these outliers are haptically irrelevant. The slight spatial difference of haptic events can hardly be felt by the user comparing two independent puncture attempts along the same path. For example, the slightly displaced detection of the skin or differently detected tissues around the fascial layer, causes out of sync force emission with potentially high MAE errors (max. 1.6 N) in the pre-puncture phase of the haptic algorithm near to the fascial layer (p. 10). The segmentation of the surrounding tissue of the fascial layers can be slightly different, causing misaligned force incline start levels of the pre-puncture phase on the needle trajectory. Again, more important is the correct sequence of events and accurately reaching the puncture goal. This is guaranteed by low segmentation errors and operator reviewed segmentation of the structures, i.e. especially the bile ducts. This argument is supported by currently published, experimentally determined, spatial JNDs for the hand-arm system[25, 26]. The elaborated JND thresholds are undercut despite some outliers at the liver fringe in the outlier HSD metric (Fig. 5). First, the liver is encountered directly after the very salient fascia puncture event with 2.7 N. Second, as seen from Table 1, the emitted forces in the case of the liver fringe are not exceptionally salient. This is especially true of the surface penetration threshold $T_N$ of the liver, which is the lowest of all tissues. Therefore, we can accept force errors stemming from seldom spatial liver fringe mismatches at this juncture.

| Tissue | Tissue Type | Reference Repres. | Modeling Method | New Repres. | Lower Interval Thresh. [HU] | Haptic Params. | | |
|---|---|---|---|---|---|---|---|---|
| | | | | | | $T_N[N]$ | $R[N]$ | $k[N/mm]$ |
| Air | Pass/Risk* | Mask | HMTF | Interval | −1024 | 0 | 0 | 0 |
| Skin | Pass | Mask | HMTF | Interval | $t_0$ | 0.7 | 0.7 | 0.8 |
| Fat/soft t. | Pass | Mask | HMTF | Interval | $t_1$ | 0.7 | 0.7 | 1.0 |
| Bone (sensitive) | Risk | Mask | HMTF | Interval | $t_2^-(\mathbf{x})$* | ∞ | 3.0 | 2.0 |
| Bone (specific) | Risk | Mask | HMTF | Interval | $t_2^+(\mathbf{x})$* | ∞ | 3.0 | 2.0 |
| Fascia | Pass | Mask | 3DS | Mask | N/A | 2.5 | 1.0 | 1.0 |
| Liver | Pass | Mask | MAS | Mask | N/A | 0.3 | 0.9 | 1.2 |
| Hep. blood | Risk | Mask | VF | Mask | N/A | 1.05 | 0.75 | 1.1 |
| Hep. bile | Target | Mask | VF | Mask | N/A | 1.2 | 0.5 | 1.0 |

**Table 1.** Model representation and haptic parameters used for the main tissues. Acronyms: HMTF = Histogram Matched Transfer Function, 3DS = 3D-Spline-Surface, MAS = Multi-Atlas Segmentation, VF = Vesselness Filtering. *Needle tip location **x** dependent. Note, new patient image data must be adapted to these thresholds which are dependent on the chosen reference patient[18].

We use easily implementable objective force signal metrics and relate them to the currently published spatial and force JNDs[25, 27] of the hand-arm system for haptic device usage. The JNDs are subject to ongoing research and therefore may change. The used metrics ensure that our results can still be related to new JNDs published in the future. In light of the currently available JNDs for the hand-arm system from refs 25, 27, of 15–26% (0.12–0.21 N) and 11% (3 mm) for the force and distance magnitudes respectively, our mean errors are small (Figs 4 and 6). Also the high percentage of absolutely correctly rendered forces (88% of positions) underlines this statement, only 12% of all path positions are affected by errors.

Our algorithm for axial force rendering and the haptic parameters for deformable virtual patients, have not been published in this level of detail and in this scale of a study with a high number of test paths from 10 test patients before. In addition, our evaluation methodology has been clearly refined, e.g. using new per test path spatial error metrics and the incorporation of spatial and force JNDs. We thoroughly show in this study set-up, using 31,222 test paths, that the generated force output using the partially segmented patient image data is robust. It is shown by example that the force evaluation framework is a valid means to prove force output of haptic needle insertion simulations, without ethics approval and the experimental hassle of specimen to image registration during needle insertion. We completely circumvent *in vivo* (animal) or *in vitro* (cadaver) experiments. For our haptics research, we only use already available CT images acquired from routine clinical data. We are grateful for the guiding work of other groups producing axial force measurements with real tissues[15, 28].

Finally, with regards to the potential time savings in patient modeling important in clinical settings, we can fully support the use the of partially segmented data combined with the non-linear spring force output. This method offers faster patient preparation by a factor of 12[18] and better force rendering as we propose a non-linear incline at tissue borders (Pre-X). Other proxy-based methods use linear force inclines for direct haptic volume rendering[29–31].

The system setup, experimental design, observed haptic errors and interpretation lines could be helpful for other haptic algorithm developers[9]. The simulation concept could easily be transferred to other liver site intervention fields in addition to PTC or PTCD, simply by replacing the target structures (with e.g. tumors).

**Conclusion.** Our extended framework and study showed, that our methods, i.e. the use of partially segmented patient image data and haptic rendering techniques featuring a non-linear spring model, are usable for virtual needle insertion simulation. The evaluated system allows the user to experience and become familiar with the visual presentation and needle insertion forces during liver needle punctures.

Furthermore, our evaluation of surgical needle insertion simulator haptics identifies, quantifies, and interprets errors, which is important for algorithm development.

In summary, the force errors of the presented force rendering algorithms using partially segmented patient data are small. In 12% of the path positions the detected errors are haptically irrelevant for the user experience. This is due to the fact that the puncture incident mismatch is mostly below JND-thresholds in terms of insertion depth[25, 26] and very similar in force amplitude to the puncture event using gold-standard label data. The force errors are also below the force JND threshold adapted to our context[27]. Bone is rendered impenetrable and would cause our high force haptic device to exert 22 $N$. The encountered MAE force outliers in this study are clearly unrelated to puncturing false-positive bone voxels, in which case we would see 22 $N$ MAE errors. Our customizable non-linear spring model gives the user a more elastic feeling of tissue borders and thus is more realistic. It can be easily adapted to *in vivo/vitro* measured realistic force curves[6] by model fitting.

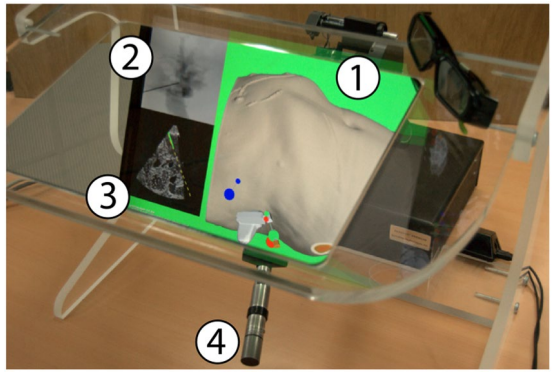

**Figure 9.** Haptic workbench and simulation: (1) Main rendering window, (2) X-ray rendering window, (3) ultrasound rendering window and (4) haptic device handle.

From a developer perspective, our evaluation framework and interpretation cues help to sort out very sensitively noticeable errors, to improve virtual patient models and haptic algorithms.

**Future Work.** Our work on force output modeling is guided by comprehensive surveys featuring *in vivo* force measurements[5, 6, 15]. In future studies, regarding axial force curve shape, we would like to investigate refined force feedback algorithms, taking into account real force measurement curves and puncture events due to different bevel tips. This way, our first haptic rendering phase could be sub-divided in two subphases: (a) initial puncture and (b) transition from tip bevel front to shaft. In reality, there are more haptic micro-events overlayed to our 4-phase curve to be included. These effects could be modeled as superposed micro processes as well. For a more granular feeling of the tissue, the CT values could be used to infer material characteristics for micro processes. The currently constant phase 3 would be more realistically modelled by a flat, non-linear incline depending on the insertion depth.

Further work will be in dynamic patient effects. More complex surgical instruments and irregular respiratory motion are currently included into the haptic simulation[10, 22, 32], work for which haptics have to be continued and evaluated accordingly. Dealing with new patient image data could be improved by researching new real-time voxel classification methods at the needle tip that for example, reliably detect the liver and its vessels. In this case, some currently necessary semi-automatic segmentation steps could become obsolete[18]. This could be achieved using high dimensional feature spaces and random forest classification[19].

## Methods

**Simulator and Virtual Patient Modeling Review and Haptics Context.** In comparison, our new VR system AcusVR-4D[9] (Fig. 9) fixes many of the short-comings of other simulators regarding the tedious and time-consuming patient-modeling (segmentation) process[1–4, 7, 11, 13]. Mainly, our system only needs a minimal subset of explicitly segmented key structures for visualization and force output. With these key structures, the system visualizes dynamic effects without the need of meshed organ models.

*Simulator and Haptics Context.* Our VR simulator features five virtual tools using proxy-based force rendering. They are controlled via the haptic device for the trainee to use in the core liver puncture workflow, i.e. virtual:

1. Palpation with a **finger tool** to search for an intercostal location for needle insertion between the 6th and 7th rib on the right side of the patient[32].
2. Inspection of the internal patient anatomy and optimization of a trajectory using the **US tool** with attached needle guide and needle line projection in the US image (Fig. 2, dashed line)[32, 33].
3. Placement of a small incision at the needle insertion point on the skin using the **scalpel tool**[32].
4. Snap-in of the needle into the needle **guide tool** at the US-device (Fig. 2).
5. Needle puncture through the incision by advancing the snapped in and axially guided needle using the **needle tool** under US imaging control (Fig. 2)[32, 33].

In step 5, the puncture is performed by guided advancement of the needle into the patient's body. Most relevantly, axial forces are emitted at the needle handle when the needle makes contact with the tissues (Fig. 9(4)). Needle guidance and axial force feedback when penetrating the patient's tissues are displayed to the user via a Geomagic 6DOF HighForce haptic device (max. 22 $N$ axial force). When a risk structure is struck by the needle, the user is notified by turning the simulation window background red, a successful insertion is indicated by a green background in the main viewport (Fig. 9(1)). Additionally, contrast agent spreading out in the bile ducts is simulated and visualized in the X-ray viewport (Fig. 9(2)). US guidance (Fig. 9(3)) effectively helps to puncture the target structure correctly and is very commonly used by physicians in this challenging puncture intervention.

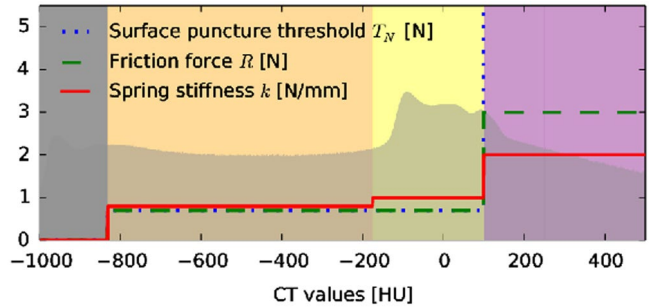

**Figure 10.** Haptic transfer function intervals over the histogram of the reference patient: color coded transfer function intervals of air (gray), skin (orange), soft-tissue (yellow) and bone (purple).

A special training mode ("scoring mode") can be used to assign scores to each training attempt along bookmarked reference paths defined by medical experts, so that the trainee can monitor his improvements during the experience with the virtual patient model.

*Virtual Patient Modeling using Partially Segmented CT data.* Here, we provide background and summarize the time-saving virtual patient modeling approach detailed in ref. 18 using explicit segmentations of relevant structures only. General pre-processing consists of removing non-body objects from CT image acquisition such as patient table, cables and medical devices attached to the patient's body. To this aim, we use a standard morphological procedure consisting of erosions, dilations, connected component analysis and hole filling. Only the biggest component, i.e. the patient's body box, is kept for further processing. Then, for new unsegmented patient images, the new intensity data is first adapted to a predefined reference patient by a histogram matching technique. Finally, regarding key structures, the fascia, the liver and its vessels are segmented using dedicated methods[18].

Our start point is a fully manually-segmented and reviewed reference patient (CT and label data) to define a general virtual patient model. For fully segmented patients, we can estimate a tissue label using Alg. 1, step 1. Each new virtual patient model then consists of the same parameter attributes inherited from the reference patient model. After the automatic intensity data adaptation, only the segmentation masks for the key structures are segmented and reviewed anew specifically for the new patient model by organ-specific, known, semi-automatic segmentation methods (Table 1).

Haptic characteristics of the fully segmented virtual patient are represented in a label-to-tuple table, which contains haptic parameters from an empirical haptics modeling phase, carried out by puncture-experienced medical experts (Table 1, right). By this means, we are able to provide haptical force feedback similar to *in vivo* force experience. Using derived parameters from *in vitro* measurements would incorporate characteristics of dead tissue, e.g. reflect fixation in formalin or non-blood perfused tissue. In our previous work on the simulation of lumbar punctures[13], medical experts were consulted to define the haptic parameter tuple ($T_N$, $R$, $k$) with cutting force threshold [$N$], friction force [$N$] and material stiffness [$N/mm$] for each label in a fully segmented patient (Table 1). For PTCD, concerning further relevant structures such as fascia, liver, blood vessels and bile ducts, we consulted again with two experienced medical experts in a tuning session to define the haptic characteristics of the new key tissues. In terms of haptic rendering, our direct force rendering method introduced first in ref. 34 uses the partially segmented patient image data.

The following patient modeling summary items consist of two paragraphs each: First, we describe the concept for the reference patient model based on a reference patient. Second, the step for its generalization to new patient models is briefly presented[18]:

---

**Algorithm 1** Case distinction to obtain the haptic parameters from a partially segmented patient in the undeformed state.

1. Check for a segmented voxel at the needle tip and use available haptic parameters.
2. If no segmentation is available, use the transfer function resulting from the tissue thresholds shown in Table 1. Switch cases depending on needle tip position **x**:
    (a) Regarding body box:
        i. Outside skin: render air with zero force.
        ii. Passed skin: if the needle encounters an air cavity inside the body, signal a risk structure.
    (b) Regarding fascia layer:
        i. Outside fascia: use sensitive bone threshold $t_2^-$.
        ii. Passed fascia: use specific bone threshold $t_2^+$.

---

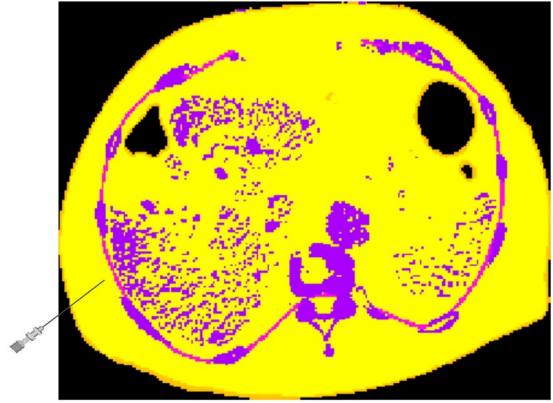

**Figure 11.** Color coded application of Fig. 10 to an undeformed axial CT slice just before puncturing the fascia.

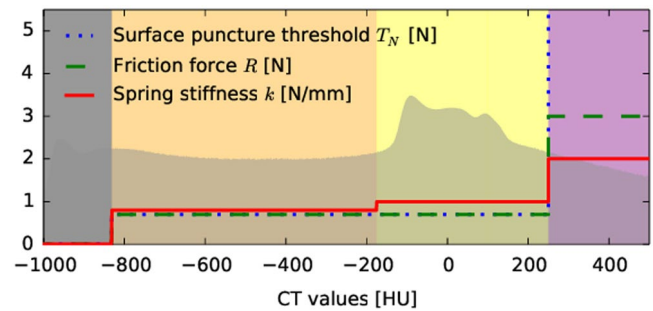

**Figure 12.** The same as Fig. 10 when the needle has passed the fascia layer (magenta).

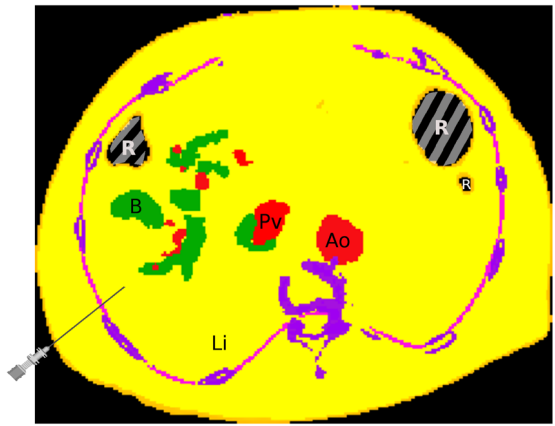

**Figure 13.** Applied transfer function from Fig. 12 augmented by some segmented key structures. The bile ducts are green and blood vessels red.

(1) Using a fully segmented reference patient, the Gaussian intensity distributions of air, skin, soft-tissue (fat) and bone are modeled. For these large volume tissue classes we determine the intersections (cf. Table 1) of the normal distribution curves $t_0$, $t_1$ and $t_2^-(\mathbf{x})$ (outside fascia) and an augmented $t_2^+(\mathbf{x})$ (inside fascia)[17, 18, 34]. A histogram matching procedure[35] is customized in ref. 18 to adapt the transfer functions and the supporting thresholds to new patient image data. From the Bayes optimal intersection points we define the air, skin, fat and bone intervals as shown by their lower thresholds in Table 1 and Figs 10, 11, 12 and 13. During simulation they are used to define the locally adaptive haptic parameter transfer functions in Fig. 10 using $t_2^-$ and Fig. 12 using $t_2^+$, respectively. This corresponds to Alg. 1, step 2(b), which differentiates between the needle tip $\mathbf{x}$ outside - resulting in Fig. 11 - and inside the fascia - resulting in Fig. 13.

(2) We fit a B-spline surface $s(\phi, h) = \left(B_x(\phi, h), B_y(\phi, h), B_z(\phi, h)\right)^{T\,36}$ through a thresholded rib cage segmentation of the reference patient obtained using the thresholds $t_2$[18].

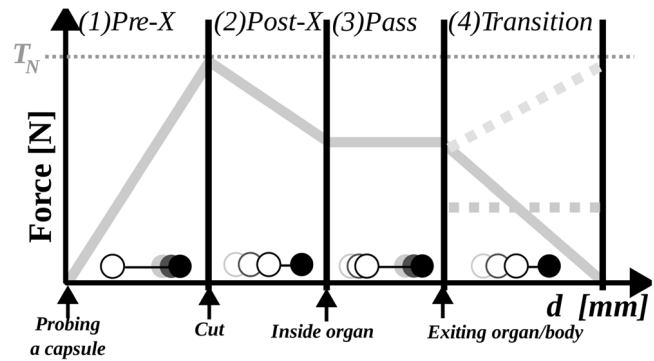

**Figure 14.** Concept of our four phases: The proxy-tip relation is indicated by circles above the x-axis (empty circles = proxy, filled circles = tip; time = gray-scale).

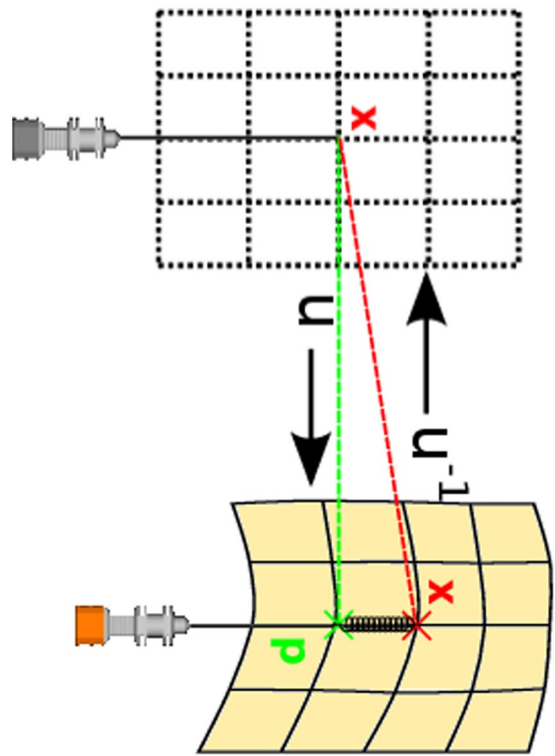

**Figure 15.** Needle in undeformed state of the virtual patient as defined by the proxy (left) and needle steered by the haptic device in the current deformed configuration of the tissue (right). Connecting **x** and **p** by a spring is used to calculate force feedback.

The fascia layer template 3D-B-spline model from the virtual reference patient is interactively adapted to the new patients rib cage by manipulating the control points. The spline model is finally converted to a label mask representation and included into the multi-label segmentation as the fascia layer of the new patient data.

(3) The liver and hepatic target structures are segmented by organ-specific approaches accompanied by typical segmentation errors. We use a multi-atlas and GraphCut segmentation concept to address the liver[18, 20, 37]. Blood vessels (risk structures to be avoided) and bile ducts (target) located inside the liver segmentation can be masked out and are ready to be addressed. We use a vessel highlighting and selection method based on multi-scale vesselness filtering[21] and auto-seeded region growing followed by morphological post-processing[18, 38].

Finally, during simulation we can estimate a label and hence the haptic parameters and force output for every voxel as shown in Alg. 1 (steps 1 and 2) in real time ($\geq$1000 Hz).

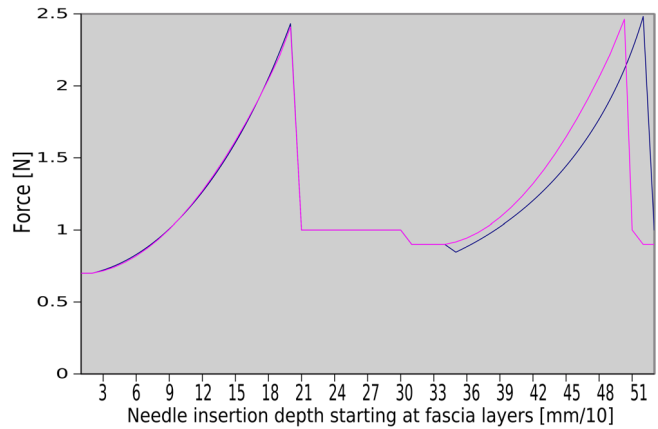

**Figure 16.** Illustrative sequence of two phase cycles from a phantom experiment with two fascial layers: Force output on the surface of organs depending on slightly different reference and segmentation (left, success mode; right, failure mode).

**Four Phase Direct Haptic Rendering with Deformable Virtual Patients.** For the axial forces at the needle tool handle, we use a proxy-based computation divided into four phases: (1) pre-puncture, (2) post-puncture, (3) pass and (4) transition (Fig. 14). In the following paragraphs, we present the used haptic algorithm in detail and review the relation to our visual deformation framework[10].

The deformation of the tissue caused by instruments or breathing, is described by a smooth (inverted) deformation field $u_t^{-1}$ at time $t$[10, 16]. Regarding tissue deformation by needle interaction we define a deformation (and force) vector $\mathbf{d} = |\mathbf{p} - \mathbf{x}|$ between two points as follows: First, the needle tip $\mathbf{x}$ lies at the same tissue related position in the undeformed and deformed spaces. Second, in the deformed state we introduce a proxy position $\mathbf{p}$ to correspond to the needle tip $\mathbf{x} = u_t^{-1}(\mathbf{p})$ in the undeformed state of the virtual patient and to a more peripheral position behind the tip as illustrated in Fig. 15 (right). Now, this vector fulfills two purposes: It is introduced as a boundary condition into the deformation field and then is smoothed to yield $u_t^{-1}$[16]. At the same time it is used as a force vector for a spring mounted between $\mathbf{x}$ and $\mathbf{p}$. We determine the haptic parameters at the proxy $\mathbf{p}$ as we follow our assumption that the spatial relation of tip and proxy defines the deformation of the tissue[9]. Thus, retrieving the parameters at the proxy position in the undeformed image data ($\mathbf{x}$), is equivalent to obtaining the parameters in a deformed state of the image ($\mathbf{p}$). Our visual deformation algorithms are described elsewhere[10, 16, 32]. In this paper, the focus lies on the details of the axial force rendering algorithms, which work in the undeformed state of the image.

Depending on the insertion depths $d = |\mathbf{d}|$ according to Hooke's law, the simplest spring model has a linear characteristic when tip and proxy diverge at a tissue border. Another variant for computation of needle forces is a non-linear spring force. Facing organ surfaces we use a second degree polynomial[15]:

$$f(d) = a_2 \cdot d^2 + a_1 \cdot d + a_0 \tag{1}$$

with $d$ being the displacement from the undeformed surface. With Eq. (1) we can use a linear spring model ($a_0 = R$, $a_1 = k$, $a_2 = 0$) as proposed in our previous work[13]. Congruent to this spring and to preserve the main haptic characteristics, the start force level $a_0$, the pre-puncture phase design parameter $a_1 \in [0, k]$ and consequently $a_2 = k \cdot (k - a_1)/|T_N - a_0|$ yield a non-linear spring, which punctures the surface at the same displacement as the linear spring to achieve similar behavior to our evaluated reference system[9, 13]. However, when the underlying gold standard and our segmentation differ only in detail, certain force errors are caused (see Fig. 16, left vs. right force incline) as force rendering is then out of sync.

---

**Algorithm 2** Haptic algorithm phases during needle insertion.

1. Pre-Puncture: the proxy is held back at a just encountered tissue surface while the actual tip advances deeper into the (deformed) tissue.
2. Postpuncture: the tissue maximum surface threshold is reached and the proxy jumps forward in the direction of the trajectory towards the tip.
3. Pass: the proxy and tip move in constant distance through the inner structure of the organ.
4. Transition: The forwarded tip encounters another organ or tissue.

---

Using Table 1 and Alg. 1 we can look up every label as $\hat{L}(\mathbf{x})$ and the corresponding haptic parameters ($T_N(\mathbf{p})$, $R(\mathbf{p})$, $k(\mathbf{p})$) at every proxy position $\mathbf{p}$ in the deformed patient data.

Our force output algorithm (see Fig. 14 and Alg. 2) works in four phases:

*Pre-puncture.* A tissue surface position $\mathbf{p}_{ph1}$ on a needle trajectory $\mathbf{N}$ can be expressed using the gradient of the augmented label data $\hat{L}$:

$$\mathbf{p}_{ph1} \in \{\mathbf{x} | \nabla \hat{L}(\mathbf{x}) \neq 0 \wedge \mathbf{x} \in \mathbf{N}\}. \quad (2)$$

In our proxy based approach[31], the virtual needle tip $\mathbf{x}$ is connected to the proxy $\mathbf{p}$ via a spring emitting force $f$ (Eq. 1). The haptic device then renders the axial non-linear spring force vector with magnitude from Eq. 1 to the user:

$$\mathbf{f}_{ph1} = f(|\mathbf{p}_{ph1} - \mathbf{x}|) \cdot \frac{\mathbf{p}_{ph1} - \mathbf{x}}{|\mathbf{p}_{ph1} - \mathbf{x}|}. \quad (3)$$

In this pre-puncture phase, by manipulation of the position of the tip behind the tissue border, 3D-forces are exerted. Simulation of the resistance of tissue surfaces to needle puncture is achieved by fixing the proxy $\mathbf{p}$ on the tissue's surface position $\mathbf{p}_{ph1}$ until the spring generates a force:

$$|\mathbf{f}_{ph1}| > T_N(\mathbf{p}_{ph1}) \quad (4)$$

higher than the tissue specific threshold. Then the cut event occurs and we enter phase 2 for a short instant in time. There the tip $\mathbf{x}$ (in the undeformed image) is already located in the internal organ structures behind the organ membrane.

*Post-puncture.* We decided to model this phase such that after the cut instance, the proxy is released from being held back at $\mathbf{p}_{ph1}$ and jumps into the inner organ structures towards $\mathbf{x}$ ($R \leq T_N$):

$$\mathbf{p}_{ph2} = \mathbf{x} + \frac{R(\mathbf{x})}{k(\mathbf{x})} \cdot \frac{\mathbf{p}_{ph1} - \mathbf{x}}{|\mathbf{p}_{ph1} - \mathbf{x}|}. \quad (5)$$

As the proxy is a virtual object only, this motivates the quick decay of the output force to the force sustain level $R$ maintained in phase 3. This provides the user with a salient feeling of cutting the membrane of an organ.

*Pass.* Advancing continuously inside an organ the modeling of viscosity comes into play. Here, we use the stiffness parameter $k(\mathbf{p})$ [N/mm] and the friction force $R(\mathbf{p})$ [N] (threshold) to define a maximum length for a spring moving behind the tip $\mathbf{x}$ that mimics advancing the needle through a viscous material:

$$l_{max} = \frac{R(\mathbf{p}_{ph3})}{k(\mathbf{p}_{ph3})}. \quad (6)$$

In contrast to the membrane puncturing, a standard linear Hook formulation is used to render the forces of the moving spring:

$$\mathbf{f}_{ph3} = k(\mathbf{p}_{ph3}) \cdot \frac{\mathbf{p}_{ph3} - \mathbf{x}}{|\mathbf{p}_{ph3} - \mathbf{x}|}. \quad (7)$$

The force emission could be improved by modelling $k$ as a function with additional dependence on a displacement parameter, however, this is left to a future work. The spring movement is implemented by moving the proxy $\mathbf{p}_{ph3}$, if the following condition is violated: $|\mathbf{f}_{ph3}| \geq R$ or rather $|\mathbf{p}_{ph3} - \mathbf{x}| \geq l_{max}$:

$$\mathbf{p}_{ph3} = \mathbf{x} + l_{max} \cdot \frac{\mathbf{p}_{ph3} - \mathbf{x}}{|\mathbf{p}_{ph3} - \mathbf{x}|} \quad (8)$$

Thus the distance between proxy and spring is always $l_{max}$ at maximum. This way when advancing or retracting the needle, the proxy moves in constant distance behind or before the tip, resulting in a ductile feeling.

*Transition.* This phase serves as a link between cycles or exit, depending on the haptic parameters of a newly encountered tissue. Exiting an organ means to encounter a border again. When the force $R$ from phase 3 is greater than the new structure's shell force threshold $T_N$, the surface is cut immediately and we skip phase 1 of the new material. This results in a decline of the force as no cutting forces are due at this instant. This case is implemented as force drop to $\mathbf{f}_{ph2}$ (phase 2). If the needle tip meets a harder tissue ($T_N > R$), we start with phase 1 of the new material resulting in a force incline. The trivial but uncommon case in our scenario is exiting the body with a force decline towards 0 (piercing).

Regarding needle retraction, only phase 3 with inverse directions is used, as all membranes have already been cut along the needle path. Some tissues are treated in a special way in our context: For bone, a special rule applies, it is rendered as impenetrable. Encountering bone at the tip emits maximum force. As in our intervention scenario, the physician introduces the needle through a small incision, for skin the sustained force level $R$ is the same as the cut force level $T_N$ (0.7 N) (Table 1).

**Force Feedback Evaluation.** We establish ground truth on the 10 CT test data sets spanning the lower thorax and upper abdomen. The spacing between the voxels in the data sets is between 0.775–0.925 mm in the x and y directions and 0.825–1 mm in z direction. CT images were acquired in portal venous phase. The necessary full volume manual expert segmentations used as gold-standard were carried out by a team of three experts.

For axial force output assessment we measure and compare forces along preplanned test paths hitting the target. Instead of curved insertion paths, we use straight paths planned in the undeformed reference space (Fig. 15, left). There, needle to target alignment is possible by finding straight lines of sight, which can then be used for trajectory planning from skin to target. Generally in this paper, we investigate force errors stemming from axial structure border distance errors. We compare reference and new algorithm 1D axial force signals acquired on preplanned trajectories, with common signal distance metrics such as root mean squared errors and maximum absolute error. There are three steps outlined in the following sections:

1. Trajectory ("reference path") planning specific to the gold-standard virtual patient. This step produces paths in proximity to the usual access area between the 6th and 7th rib on the right side of the patient,
2. Simulated robotic needle steering to accurately follow the trajectories and 1D force signal sampling on these paths using a reference (fully manual segmentation) and a new algorithm (transfer functions augmented by partial segmentation),
3. Structure border lag analysis, force signal comparison and statistical analysis.

Extensive quantitative evaluations take place for force output of the haptic algorithm. We use many target (center line of bile ducts) hitting paths from a surgery planning step for quantitative evaluation of force feedback. Automatic needle steering along approximately 3000 paths from each of the ten test patients ("reference paths") yields reproducible force outputs for the algorithms to be compared in ten patients (see Fig. 3). We use occlusion, distance from critical structures and penetration depth into bile vessels as quality criteria.

*Path Planning and Needle Steering.* Visibility checks serve to generate a high number of test paths. Ray casting on the GPU is used for planning[17]. Trajectories affected by hard constraints are removed from the candidate path set. Acceptable paths reach the bile target within a maximum needle insertion length of 90 mm. In our path quality scoring, we use a concept similar to Baegert *et al.*[39] adapted to our needs. The hard constraints for path dismissal comprise (1) visibility of a target structure voxel from the skin and (2) needle insertion length (9 cm). Individually rated soft constraints yield a path score in a weighted averaging manner.

We use the following criteria for candidate paths starting from every skin voxel:

- $C_1$: The visibility hard constraint dismisses all trajectories that are blocked on their way from the skin to the target through a risk structure such as bones or blood vessels. Failure is indicated by a value of infinity, success by 0.
- $C_2$: Distance to target. The target must be reached in 9 cm path length. Failure is indicated by a value of infinity, success by 0.
- $C_3$: The minimal distance to critical structures soft constraint scores a trajectory by calculation of the smallest distance to a risk structure on the path.
- $C_4$: The target hit soft criterion counts the voxels along the needle shaft. This counting is conducted inside a cut target structure (bile duct) only and tests for a certain number of voxels being visited. A higher number indicates a favorable penetration depth and insertion angle in line with a bile duct.

The final score for a path **N** is a normalized average of the criterion scores:

$$Q_{\mathbf{N}} = \sum_{i=1}^{4} \alpha_i \cdot C_i(\mathbf{N}). \tag{9}$$

The criterion scores $C_{3,4}$ are normalized to 1. Weights $\alpha_i = \frac{1}{2}$ determine the influence of the criteria. Paths with $Q_{\mathbf{N}} = \infty$ are dismissed right away. For each skin voxel we keep the best path from the candidate set as reference path.

The following rationale supports the criteria set-up. In treatment, it is not advised to puncture through and closely below the rib bones (risk structure), where fine nerves and vessels are located causing pain if struck. Blood vessels should also not be close to the needle. Hence, nearby bone and blood vessel paths are removed from the candidate set by a quality threshold, i.e. $Q_{\mathbf{N}} < 0.4$. Also, falsely detected bone voxels from the transfer function classification could occur in very close proximity to the ribs, e.g. due to partial volume effects. These medical and technical arguments motivate the removal of these low score paths from the candidate set (cf. Results, Fig. 3).

To sum up, we test our simulator for successful trainee experience, i.e. starting from the skin paths hit the target with respect to the hard and quality soft constraints. Paths from the skin to the target (bile ducts) colliding with or aligned in close proximity to risk structures are dismissed. Soft constraints are used to score the candidate path set and to reduce it further.

Finally, we test our haptics on 10 patients with a total of 31,222 test paths, i.e. ca. 3.000 paths per patient.

*Quantitative Measures.* The selected paths from ten test patients are now used for force output evaluation. While steering the needle along the paths, two forces are calculated and compared. We compare forces based on the fully manually vs. partially segmented data[18] using the same non-linear spring model.

Using about 1,000 force pair samples per planned path for insertion, the worse case for sample spacing of the currently steered path can be given as 0.09 mm, which is sub-voxel resolution. Thus, generous oversampling of $\geq 10\times$ for a full length path of 90 mm (worst case) is used.

The metrics "root mean squared error" (RMSE) and "maximum absolute error" (MAE) are used to assess the axial force differences from the experiment. We use (1) mean and standard deviation $\sigma$, (2) account for top force outliers higher than $2.7 \cdot \sigma$ and (3) use box plots in terms of MAE to better show systematic errors. The analysis of the higher MAE outliers can be important for method investigations during haptic algorithm development. For the evaluation of the segmentation errors caused by out of sync force rendering (Fig. 16) the mean surface distance (MSD) and Hausdorff distance (HSD) are used per path for the structures hit by the needle. We relate the errors to just noticeable differences (JNDs) for length and force for the hand-arm system[25–27].

The so-called Weber fraction given in reference force or distance magnitude % with a perceived reference stimulus $S$ and a just perceivable relative stimulus change $\Delta S$[27]:

$$JND\% = \frac{(S + \Delta S) - S}{S} \cdot 100 \qquad (10)$$

is used to quantify JNDs. Here we regard forces and distances as stimuli. We consider 2–3 mm as an acceptable just noticeable spatial distance error range for salient needle force events to occur based on reasonable assumptions[25, 26]. Note, that empirically measured JNDs for distances with the finger span method[26] are twice as low as for the upper limb movements[25]. The upper limb movement is more relevant in our context. A 2 mm lower threshold corresponds to a conservative JND of 8% distance related to an average reference length of 28 mm. This corresponds to the most important sub-distances for our needle to bypass, i.e. the liver to bile duct border and skin to fascia border path segments. The distance of 3 mm corresponds to a JND of 11% distance. This value is justified by ref. 25 and corresponds to a doubling of the JND from the finger-span method for 28 mm[26]. Generally, JNDs for decreasing reference measures are higher above average in small scale domains.

Hand-arm system force JNDs range from 15 to 26% force magnitude for high (6 N) to low (0.5 N) reference forces[27]. Regarding those human force JNDs we relate a hand-arm JND of 18.1% force magnitude (for $-0.87$ N, see Table II in ref. 27) to the average counter force inside the body during insertion of 0.8 N, yielding a threshold of 0.145 N.

To further assess the path position fraction where the segmentation errors cause force errors, we also asses the strict measure of the percentage of exactly identical emitted forces for every path $\mathbf{N}$:

$$\%F_c = \frac{|\{\mathbf{x} | f_{ref}(\mathbf{x}) - f_{test}(\mathbf{x}) = 0, \mathbf{x} \in \mathbf{N}\}|}{|\{\mathbf{x} | \mathbf{x} \in \mathbf{N}\}|} \cdot 100. \qquad (11)$$

In summary, in the conducted experiment the segmentation fundament is varied: For the reference forces we use the full manual expert segmentation, for the test forces the transfer functions augmented by partially segmented data are used. Using these metrics and study design, we mainly aim to evaluate the influence of the segmentation errors caused by our patient modeling strategy on the force rendering in our haptic simulation.

## References

1. Ullrich, S. & Kuhlen, T. Haptic Palpation for Medical Simulation in Virtual Environments. *IEEE Trans Vis Comput Graphics* **18**, 617–25, doi:10.1109/TVCG.2012.46 (2012).
2. Coles, T., John, N., Gould, D. & Caldwell, D. Integrating Haptics with Augmented Reality in a Femoral Palpation and Needle Insertion Training Simulation. *IEEE Trans Hapt* **4**, 199–209, doi:10.1109/TOH.2011.32 (2011).
3. Villard, P. F. *et al*. Interventional Radiology Virtual Simulator for Liver Biopsy. *Int J Comp Ass Rad Surg* **9**, 255–267, doi:10.1007/s11548-013-0929-0 (2014).
4. Ni, D., Chan, W., Qin, J. & Chui, Y. A Virtual Reality Simulator for Ultrasound-guided Biopsy Training. *IEEE Compu Graph Appl* **11**, 143–150, doi:10.1109/MCG.2009.151 (2011).
5. Abolhassani, N., Patel, R. & Moallem, M. Needle Insertion into Soft Tissue: A Survey. *Med Eng Phys* **29**, 413–31, doi:10.1016/j.medengphy.2006.07.003 (2007).
6. Gerwen, D. Jv, Dankelman, J. & Dobbelsteen, J. Jvd Needle-tissue Interaction Forces–a Survey of Experimental Data. *Med Eng Phys* **34**, 665–80, doi:10.1016/j.medengphy.2012.04.007 (2012).
7. Willaert, W. I. M., Aggarwal, R., Van Herzeele, I., Cheshire, N. J. & Vermassen, F. E. Recent Advancements in Medical Simulation: Patient-specific Virtual Reality Simulation. *World J Surg* **36**, 1703–12, doi:10.1007/s00268-012-1489-0 (2012).
8. Duriez, C., Guebert, C., Marchal, M., Cotin, S. & Grisoni, L. Interactive Simulation of Flexible Needle Insertions Based on Constraint Models. *Proc. MICCAI* **5762**, 291–299, doi:10.1007/978-3-642-04271-3_36 (2009).
9. Fortmeier, D., Mastmeyer, A., Schröder, J. & Handels, H. A Virtual Reality System for PTCD Simulation using Direct Visuo-haptic Rendering of Partially Segmented Image Data. *IEEE J Biomed Health Inform* **20**, 355–366, doi:10.1109/JBHI.2014.2381772 (2016).
10. Fortmeier, D., Wilms, M., Mastmeyer, A. & Handels, H. Direct Visuo-haptic 4D Volume Rendering using Respiratory Motion Models. *IEEE Trans Hapt* **8**, 371–383, doi:10.1109/TOH.2015.2445768 (2015).
11. Villard, P.-F. *et al*. A Prototype Percutaneous Transhepatic Cholangiography Training Simulator with Real-time Breathing Motion. *Int J Comp Ass Rad Surg* **4**, 571–578, doi:10.1007/s11548-009-0367-1 (2009).
12. Liwu, L. Practical Clinical Ultrasound Diagnosis. *World Scientific Publishing Company*, ISBN: 9789810229221 (1997).
13. Färber, M., Hummel, F., Gerloff, C. & Handels, H. Virtual Reality Simulator for the Training of Lumbar Punctures. *Methods Inform Med* **48**, 493–501, doi:10.3414/ME0566 (2009).
14. Wu, Y. & Yencharis, L. Commercial 3D Imaging Software Migrates to PC Medical Diagnostics. *Advanced Imaging Magazine*, 16–21 (1998).
15. Okamura, A. M., Simone, C. & O'Leary, M. D. Force Modeling for Needle Insertion into Soft Tissue. *IEEE Trans Biomed Eng* **51**, 1707–1716, doi:10.1109/TBME.2004.831542 (2004).
16. Fortmeier, D., Mastmeyer, A. & Handels, H. Image-based Soft Tissue Deformation Algorithms for Real-time Simulation of Liver Puncture. *Current Medical Imaging Reviews* **9**, 154–165, doi:10.2174/1573405611309020011 (2013).
17. Mastmeyer, A., Hecht, T., Fortmeier, D. & Handels, H. Ray-casting based Evaluation Framework for Haptic Force Feedback during Percutaneous Transhepatic Catheter Drainage Punctures. *Int J Comp Ass Rad Surg* **9**, 421–431, doi:10.1007/s11548-013-0959-7 (2014).
18. Mastmeyer, A., Fortmeier, D. & Handels, H. Efficient Patient Modeling for Visuo-haptic VR Simulation using a Generic Patient Atlas. *Comp Meth Prog Bio* **132**, 161–175, doi:10.1016/j.cmpb.2016.04.017 (2016).

19. Mastmeyer, A., Fortmeier, D., Mastmeyer, A. & Handels, H. Random Forest Classification of Large Volume Structures for Visuo-haptic Rendering in CT Images. *Proc. SPIE* **9784**, 97842H–97842H–8, doi:10.1117/12.2216845 (2016).
20. Mastmeyer, A., Fortmeier, D., Maghsoudi, E., Simon, M. & Handels, H. Patch-based Label Fusion using Local Confidence-measures and Weak Segmentations. *Proc. SPIE* **8669**, 86691N–1–86691N–11, doi:10.1117/12.2006082 (2013).
21. Sato, Y. *et al.* Three-dimensional multi-scale Line Filter for Segmentation and Visualization of Curvilinear Structures in Medical Images. *Med Image Anal* **2**, 143–168, doi:10.1016/S1361-8415(98)80009-1 (1998).
22. Fortmeier, D., Mastmeyer, A. & Handels, H. Image-Based Palpation Simulation With Soft Tissue Deformations Using Chainmail on the GPU. *German Conference on Medical Image Computing (BVM)* **2013**, 140–145, doi:10.1007/978-3-642-36480-8_26 (2013).
23. Fortmeier, D., Mastmeyer, A. & Handels, H. Optimized Image-Based Soft Tissue Deformation Algorithms for Visualization of Haptic Needle Insertion. *Stud Health Technol Inform* **184**, 136–140 (2013).
24. Fortmeier, D., Mastmeyer, A. & Handels, H. GPU-based Visualization of Deformable Volumetric Soft-Tissue for Real-time Simulation of Haptic Needle Insertion. *German Conference on Medical Image Computing (BVM)* **2012**, 117–122, doi:10.1007/978-3-642-28502-8 (2012).
25. van Beek, F. E., Bergmann Tiest, W. M. & Kappers, A. M. L. Haptic discrimination of distance. *PLoS ONE* **9**, 1–9, doi:10.1371/journal.pone.0104769 (2014).
26. Tan, H. Z., Pang, X. D. & Durlach, N. I. Manual resolution of length, force, and compliance. *Advances in Robotics* **42**, 13–18 (1992).
27. Vicentini, M., Galvan, S., Botturi, D. & Fiorini, P. Evaluation of Force and Torque Magnitude Discrimination Thresholds on the Human Hand-arm System. *ACM Trans Appl Percept* **8**, 1:1–1:16, doi:10.1145/1857893 (2010).
28. Majewicz, A. *et al.* Behavior of tip-steerable needles in *ex vivo* and *in vivo* tissue. *IEEE Trans Biomed Eng* **59**, 2705–2715, doi:10.1109/TBME.2012.2204749 (2012).
29. Lundin, K., Gudmundsson, B. & Ynnerman, A. General Proxy-Based Haptics for Volume Visualization. *Proc. Eurohaptics* **2005**, 557–560, doi:10.1109/WHC.2005.62 (2005).
30. Lundin, K., Ynnerman, A. & Gudmundsson, B. Proxy-based haptic feedback from volumetric density data. *Proc. Eurohaptics* **2002**, 104–109 (2002).
31. Ruspini, D. C., Kolarov, K. & Khatib, O. The Haptic Display of Complex Graphical Environments. *Proc. Computer Graphics and Interactive Techniques (SIGGRAPH)* **1997**, 345–352, doi:10.1145/258734 (1997).
32. Fortmeier, D., Mastmeyer, A. & Handels, H. An Image-Based Multiproxy Palpation Algorithm for Patient-Specific VR-Simulation. *Medicine Meets Virtual Reality 21 (MMVR), Stud Health Technol Inform* **196**, 107–113, doi:10.3233/978-1-61499-375-9-107 (2014).
33. Mastmeyer, A., Wilms, M., Fortmeier, D., Schröder, J. & Handels, H. Real-time Ultrasound Simulation for Training of US-guided Needle Insertion in Breathing Virtual Patients. *Medicine Meets Virtual Reality 22 (MMVR), Stud Health Technol Inform* **220**, 219–226, doi:10.3233/978-1-61499-625-5-219 (2016).
34. Mastmeyer, A., Fortmeier, D. & Handels, H. Direct Haptic Volume Rendering in Lumbar Puncture Simulation. *Medicine Meets Virtual Reality 19 (MMVR), Stud Health Technol Inform* **173**, 280–286, doi:10.3233/978-1-61499-022-2-280 (2012).
35. Nyúl, L., Udupa, J. K. & Zhang, X. New Variants of a Method of MRI Scale Standardization. *IEEE Trans Med Imag* **19**, 143–150, doi:10.1109/42.836373 (2000).
36. Dierckx, P. Curve and Surface Fitting with Splines. *Oxford University Press*, ISBN: 9780198534402 (1995).
37. Stawiaski, J., Decenciere, E. & Bidault, F. Interactive Liver Tumor Segmentation using Graph Cuts and Watershed. *Medical Image Computing and Computer-Assisted Intervention (MICCAI) - Workshop on 3D Segmentation in the Clinic: A Grand Challenge II. Liver Tumor Segmentation* (2008).
38. Adams, R. & Bischof, L. Seeded Region Growing. *IEEE Trans Pattern Anal Mach Intell* **16**, 641–647, doi:10.1109/34.295913 (1994).
39. Baegert, C., Villard, C., Schreck, P. & Soler, L. Multi-criteria Trajectory Planning for Hepatic Radiofrequency Ablation. *Proc. MICCAI* **2007**, 676–684, doi:10.1007/978-3-540-75759-7_82 (2007).

### Acknowledgements

This work is supported by grant DFG HA 2355/11-2. The authors would like to thank Hala El-Shaffey for proofreading.

### Author Contributions

A.M. developed algorithms, conceived, conducted the experiments and analyzed the results, D.F. developed algorithms. A.M., D.F., H.H. reviewed the manuscript.

### Additional Information

**Competing Interests:** The authors declare that they have no competing interests.

**Publisher's note:** Springer Nature remains neutral with regard to jurisdictional claims in published maps and institutional affiliations.